\documentclass[reprint,aip,onecolumn]{revtex4-2}

\usepackage{graphicx}
\usepackage{nccmath}
\usepackage{epstopdf, epsfig}
\usepackage{amsmath, amssymb}
\usepackage{bm}
\usepackage{here}
\usepackage{graphicx,siunitx}
\usepackage{listings,color,courier,natbib}
\graphicspath{{}}
\usepackage[abs]{overpic}

\newcommand{\fg}[1]{{{#1}}}
\draft 

\begin{document}

\title{Convolutional neural network based hierarchical autoencoder for nonlinear mode decomposition of fluid field data}

\author{Kai Fukami}
\email[]{kai.fukami@keio.jp}
\affiliation{Department of Mechanical Engineering, Keio University, Yokohama 223-8522, Japan}

\author{Taichi Nakamura}
\email[]{taichi.nakamura@kflab.jp}
\affiliation{Department of Mechanical Engineering, Keio University, Yokohama 223-8522, Japan}

\author{Koji Fukagata}
\email[]{fukagata@mech.keio.ac.jp}
\affiliation{Department of Mechanical Engineering, Keio University, Yokohama 223-8522, Japan}

\date{\today}

\begin{abstract}
We propose a customized convolutional neural network based autoencoder called a hierarchical autoencoder, which allows us to extract nonlinear autoencoder modes of flow fields while preserving the contribution order of the latent vectors.
As preliminary tests, the proposed method is first applied to a cylinder wake at ${\rm Re}_D=100$ and its transient process.
It is found that the proposed method can extract the features of these laminar flow fields as the latent vectors while keeping the order of their energy content.
The present hierarchical autoencoder is further assessed with a two-dimensional $y-z$ cross-sectional velocity field of turbulent channel flow at ${\rm Re}_\tau=180$ in order to examine its applicability to turbulent flows.
It is demonstrated that the turbulent flow field can be efficiently mapped into the latent space by utilizing the hierarchical model with a concept of {\it ordered autoencoder mode family}.
The present results suggest that the proposed concept can be extended to meet various demands in fluid dynamics including reduced order modeling and its combination with linear theory-based methods by using its ability to arrange the order of the extracted nonlinear modes.

\end{abstract}

\pacs{}

\maketitle 
\section{Introduction}

In the fluid dynamics community, various reduced order modeling techniques have been utilized as prominent tools not only to analyze nonlinear phenomena with chaotic nature but also to design efficient flow control schemes.
Especially among them, linear theory-based mode decomposition methods such as proper orthogonal decomposition (POD) \citep{Lumely1967} and dynamic mode decomposition (DMD) \citep{Schmid2010} have been widely utilized to extract the dominant spatio-temporal coherent structures in flow fields \citep{TBDRCMSGTU2017,THBSDBDY2019}.
For instance, POD modes, which have an orthogonality with each other, can be used to construct the Galerkin projection-based reduced order models \citep{NAMTT2003,NPM2005};  
DMD is also used as the post processing tool for flow analyses in both numerical and experimental studies \citep{LANOT2018,Schmid2011}.
In this way, the reduced order modeling has played a significant role to understand the high-dimensional nonlinear complex flow phenomena by mapping them into low-dimensional systems.

{In addition to the aforementioned efforts based on the linear methods, non-intrusive reduced order models aided by machine learning have shown promising results for several fluid flow applications \citep{pawar2020data,pawar2019deep,renganathan2020machine,deng2019time}.}
{In particular, unsupervised machine learning based reduced order modeling with} autoencoders (AEs) \citep{HS2006} has emerged so as to account for the nonlinearity in the low-dimensional mapping via activation functions \citep{BNK2020,BEF2019,THU2020,LKB2018,NG2020,BHT2020}.  
To the best of our knowledge, the first attempt in the fluid dynamics community was that of \citet{Milano2002}.
They applied a multi-layer perceptron based AE to the randomly forced Burgers equation and a turbulent channel flow and demonstrated its superior reconstruction performance over the linear POD, thanks to the nonlinear mapping with nonlinear activation functions.  
More recently, combination with convolutional neural networks (CNNs) has also been suggested for effective low-dimensional mapping \citep{HFMF2020a,HFMF2020b,HFMF2019,omata2019,CLKB2019,EMM2019,XD2019,MLB2020,LPBK2020,maulik2019time}.  
\citet{MFF2019} proposed a nonlinear mode decomposition method with a convolutional neural network named the {\it mode decomposing convolutional neural network autoencoder} (MD-CNN-AE) to visualize the autoencoder modes.
They applied this MD-CNN-AE to a laminar cylinder wake and its transient process.
It was revealed that the MD-CNN-AE can map these flows into a two-dimensional latent vector containing also the higher-order POD modes; however, they also reported that some well-designed methods may be required to handle more complex flows such as turbulent flows.
This is due to the fact that a lot of spatial modes are required to represent finer spatial coherent structures of turbulence \citep{alfonsi2006,MPMF2019}.
Another crucial issue on autoencoder-based reduced order modeling is that the encoded latent modes cannot be arranged in the order of energy contributions 
unlike POD.

To overcome the aforementioned issues, we utilize the idea of the hierarchical autoencoder proposed by \citet{SSH2004} 
which can extract the modes in their contribution order for reconstruction while achieving more efficient data compression than the conventional autoencoders \citep{saegusa2005nonlinear}.  
In the present study, we extend the approach of \citet{SSH2004} to the CNN autoencoder.
The present machine learning model is first applied to a two-dimensional laminar cylinder wake at ${\rm Re}_D=100$ and its transient process.  
We then propose a concept of {\it ordered autoencoder mode family} to efficiently deal with problems of higher dimensions and assess its ability by considering $y-z$ cross-sectional velocity fields of a turbulent channel flow at ${\rm Re}_\tau =180$.  
At last, in the concluding remarks, we provide an outlook with considerable applications of the present model.

\section{Convolutional neural network based hierarchical autoencoder}\label{sec2}

\begin{figure*}
	\begin{center}
		\includegraphics[width=0.70\textwidth]{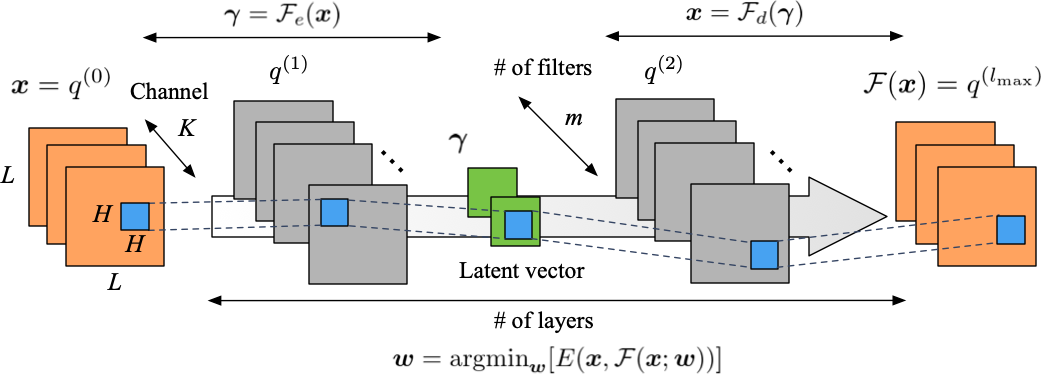}
		\caption{Two-dimensional convolutional neural network-based autoencoder with three hidden layers.  Orange squares represent the input and output vectors $\bm x$.  The latent vector $\bm \gamma$ is illustrated as the green sections.}
		\label{fig1}
	\end{center}
\end{figure*}

For constructing a hierarchical autoencoder (AE) in the present study, we use a convolutional neural network (CNN) \citep{LBBH1998} so that a high-dimensional flow field can be mapped into a low-dimensional latent space.
Let us present in figure \ref{fig1} the illustration of two-dimensional CNN-based AE with three hidden layers as an example.  
With unsupervised learning, a machine learning model $\cal F$ is trained to output the same data as the input data $\bm x$ such that ${\bm x}\approx{\cal F}({\bm x};{\bm w})$, where ${\bm w}$ denotes the weights in the machine learning model.  
Mathematically speaking, in the training process for obtaining the machine learning model $\cal F$, the weights $\bm w$ are optimized to minimize the prescribed error function $E$ such that ${\bm w}={\rm argmin}_{\bm w}[{E}({\bm x},{\mathcal F}({\bm x};{\bm w}))]$.  
A notable point here is that the dimension of the internal structure ${\mathbb R}^{\bm \gamma}$ illustrated as the green sections in figure \ref{fig1} is smaller than that of the input or output ${\mathbb R}^{\bm x}$.  
Hence, if the model has been successfully trained to output the data approximately equal to the input data, it indicates that the high-dimensional data have been successfully mapped into the low-dimensional latent vector $\bm \gamma$.  
In sum, these relations can be formulated as
\begin{align}
    {\bm{\gamma}} ={\mathcal F}_{e}({\bm x}), ~~{\bm{x}} \approx{\mathcal F}_{d}(\bm \gamma),
\end{align}
where ${\cal F}_e$ and ${\cal F}_d$ are, respectively, the encoder and decoder parts of the autoencoder, as shown in figure \ref{fig1}.
Note in passing that how much we can suppress the dimension, of course, highly depends on the nature of the flow field dealt with.
In the present study, we use the $L_2$ norm error as the error function $E$, and the Adam optimizer \citep{kingma2014} is applied for updating the weights in the iterative training process.

\begin{figure*}
	\begin{center}
		\includegraphics[width=0.5\textwidth]{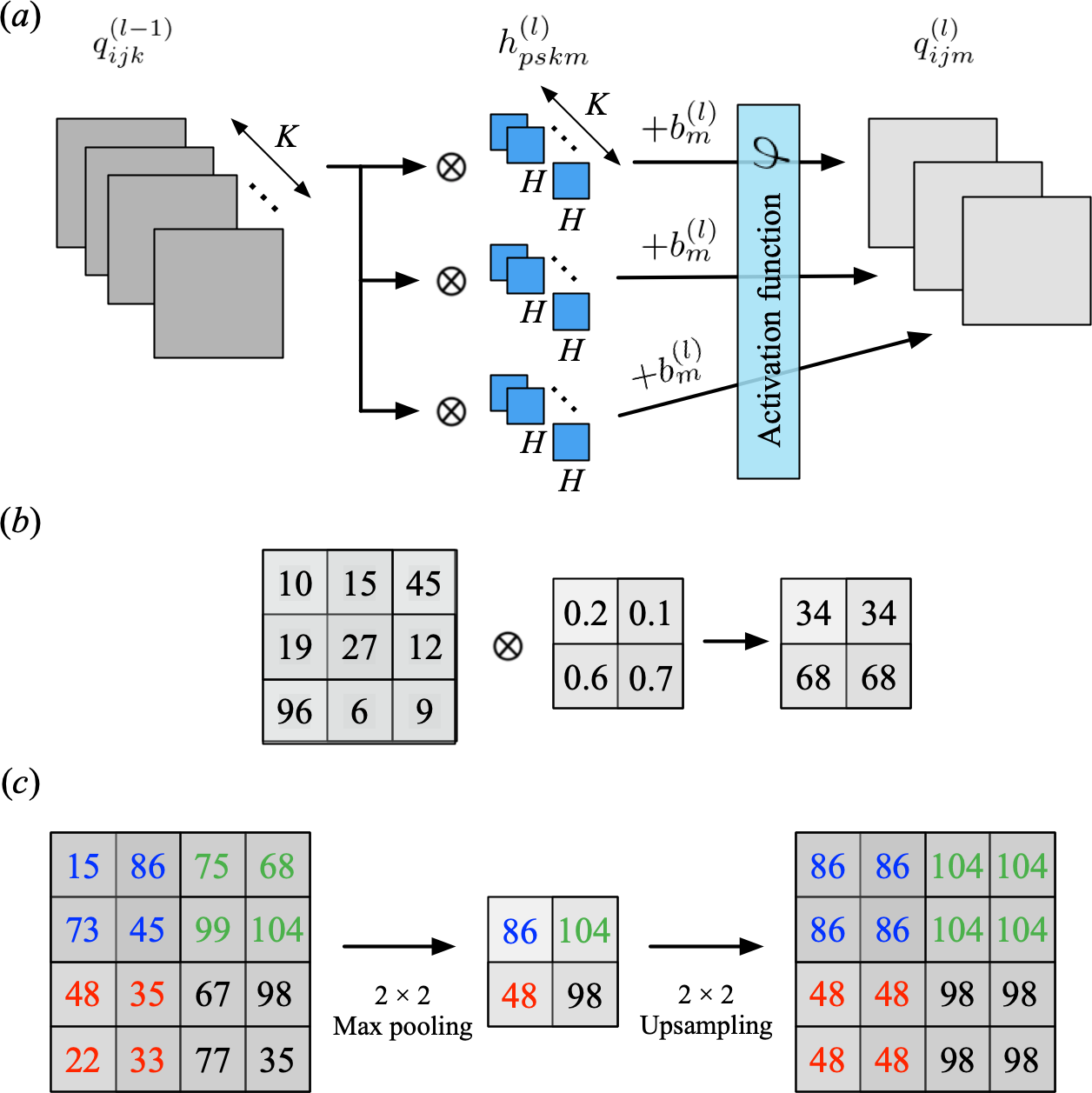}
		\caption{{Schematics of convolutional neural network: $(a)$ the structure of a single convolutional layer; $(b)$ an example of convolution operation; $(c)$ an example of pooling and upsampling operations.}}
		\label{fig1-1}
	\end{center}
\end{figure*}

The CNN is trained with the concept of weight sharing.
{As illustrated in figures \ref{fig1-1}$(a)$ and $(b)$,} the output $q^{(l)}$ of a node of CNN at layer $l$, location $(i,j)$, and filter index $m$, is obtained by convolving the filter $h^{(l)}$ (illustrated as the blue $H\times H$ squares in figure \ref{fig1}) and the output of the upstream layer $q^{(l-1)}$ as
\begin{eqnarray}
    q^{(l)}_{ijm} = {\varphi}\biggl({b_m^{(l)}}+\sum^{K-1}_{k=0}\sum^{{H}-1}_{p=0}\sum^{{H}-1}_{s=0}h_{p{s}km}^{(l)} q_{i+p{{-C}}\,j+{s{-C}}\,k}^{(l-1)}\biggr),
\end{eqnarray}
where {$C={\rm floor}(H/2)$,} $K$ is the number of \fg{filters in a convolution layer --- in the input and output layers, it corresponds to the number of flow variables per each point ---}, {$b_m^{(l)}$ is the bias,} and $\varphi$ is an activation function, which is usually a monotonically increasing nonlinear function. 
In the present paper, the hyperbolic tangent function $\varphi(s)=(e^{s}-e^{-s})\cdot(e^{s}+e^{-s})^{-1}$ is utilized as the activation function for comparing the results with those of \citet{MFF2019}, who
chose the hyperbolic tangent function after examining several kinds of nonlinear activation functions.
{--- we have checked that the similar performance 
can also be 
obtained when the ReLU function or the sigmoid function is used instead.}
In the training process of CNN, the filter coefficients $h_{p{s}km}^{(l)}$ are optimized as the weights to obtain the desired output.  
The updated filter coefficients are shared in the same network layer, as shown in figure \ref{fig1}.  
Note that the assumption on CNN is that the pixels of image far apart have no strong correlation.
{To perform a dimension reduction and extension, which are indispensable for the construction of AE, a max pooling operation is used for the encoder part and an upsampling operation is inserted for the decoder part, whose schematics can be seen in figure \ref{fig1-1}$(c)$.
Through the max pooling operation, the AE is able to reduce the dimension of input images while obtaining the robustness against rotation and translation of the images.
The upsampling operation in the decoder part copies the values of the low-dimensional maps into a high-dimensional image, i.e., the nearest neighbor interpolation.}
Although \fg{the use of CNN was} started in computer science, this has also been adopted for dealing with various high dimensional problems including fluid dynamics \citep{FNKF2019,FFT2019a,FFT2019b,FFT2020b,MFF2020,FFT2019tcfd,LY2019,KL2020,LY2019b,jin2018prediction,liu2020deep,han2019novel,deng2019super,li2019inversion,sekar2019fast,guemes2019sensing}.

\begin{figure*}
	\begin{center}
		\includegraphics[width=0.70\textwidth]{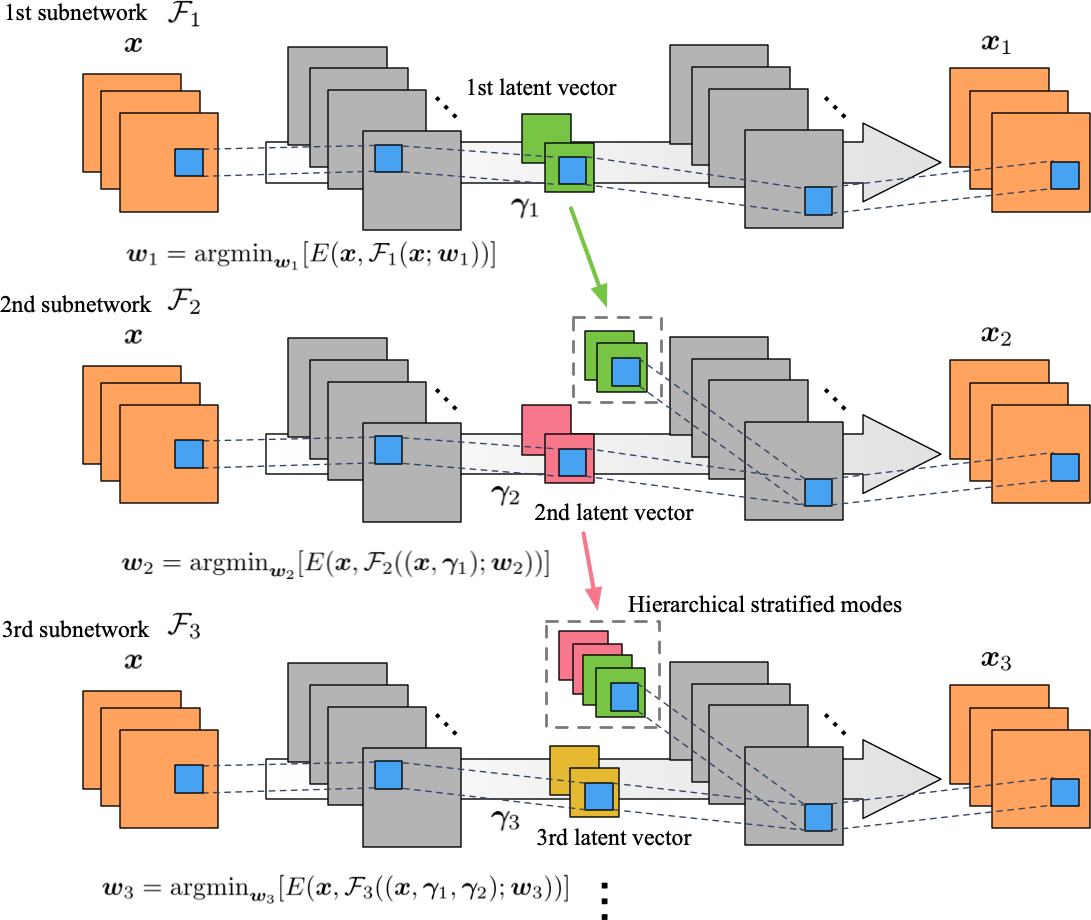}
		\caption{Two-dimensional convolutional neural network-based hierarchical autoencoder.  The first latent vector illustrated as green sections ${\bm\gamma}_1$ are stacked onto the second latent vector ${\bm \gamma}_2$, and this stacked latent vector $\{{\bm \gamma}_1,{\bm \gamma}_2\}$ is decoded by the decoder part of the second subnetwork. 
		By repeating this procedure, the ordered autoencoder modes can be obtained.}
		\label{fig2}
	\end{center}
\end{figure*}

It has been reported that AE-based low-dimensional mapping has a significant advantage against linear-theory based methods because of its structure which can take into account nonlinearities thanks to the activation functions.
However, AE-based methods usually lack interpretability: namely, it is extremely difficult to interpret the physical meanings of the latent vector extracted through nonlinear filter operations.  
More concretely, the AE-based modes have no concept like the eigenvalues or the singular values in the linear mode decomposition methods, which enables us to understand the contribution of each latent vector following their energy containing ratio. This is attributed to the fact that these AE-based modes are not orthogonal with each other.
In order to solve this issue, we apply a concept of hierarchical AE developed by \citet{SSH2004}, as illustrated in figure \ref{fig2}. 
This method can arrange the order of the latent modes based on their contributions to the reconstructed field as follows:

\begin{enumerate}
\item The first subnetwork ${\cal F}_1$ is trained to optimize the weights ${\bm w}_1$ so that we obtain the output the same as the input vector $\bm x$.  We can then obtain the first latent mode ${\bm{\gamma}}_1$ as illustrated by green squares in figure \ref{fig2}.  
The output ${\bm x}_1$ is reconstructed using the decoder part of the first subnetwork ${\cal F}_{d,1}$ with only the first mode ${\bm{\gamma}}_1$. 
\item The second subnetwork ${\cal F}_2$ is then trained to optimize the weights ${\bm w}_2$.  
Here, the first latent vector ${\bm \gamma}_1$, which has already been obtained as above, is stacked onto the second latent vector ${\bm \gamma}_2$ being updated, and this united latent vector $\{{\bm \gamma}_1,{\bm \gamma}_2\}$ is decoded by the decoder part of second subnetwork ${\cal F}_{d,2}$, as illustrated in figure \ref{fig2}.
Because the first mode ${\bm\gamma}_1$ was obtained from the individual training process, the second subnetwork ${\cal F}_2$ is led to learn the remaining features that could not be obtained in the first subnetwork ${\cal F}_1$.
\item The third subnetwork ${\cal F}_{3}$ is trained similarly to the steps 1 and 2. 
The additional inputs, i.e., the union of first and second autoencoder modes $\{{\bm \gamma}_1,{\bm \gamma}_2\}$, referred to as hierarchical stratified modes in figure \ref{fig2}, are stacked onto the third latent vector ${\bm \gamma}_3$ being updated, and this united latent vector $\{{\bm \gamma}_1,{\bm \gamma}_2, {\bm \gamma}_3\}$ is decoded by the decoder part of the third subnetwork ${\cal F}_{d,3}$,  
\end{enumerate}

By repeating these steps, we can obtain the ordered AE-based modes following their contributions to the  reconstructed field such that
\begin{eqnarray}
    {{\bm x}_1}= {\cal F}_1({\bm x};{\bm w}_1),~  {{\bm\gamma}_1}={\cal F}_{e,1}(\bm x), ~  {\bm x}_1={\cal F}_{d,1}({\bm \gamma}_1)\\
    {{\bm x}_2}= {\cal F}_2({\bm x},{{\bm \gamma}_1};{\bm w}_2),~  {{\bm\gamma}_2}={\cal F}_{e,2}(\bm x), ~  {\bm x}_2={\cal F}_{d,2}(\{{\bm \gamma}_1,{\bm \gamma}_2\})\\
    {{\bm x}_3}= {\cal F}_3({\bm x}, \{{{\bm \gamma}_1},{{\bm \gamma}_2} \};{\bm w}_3 ),~  {{\bm\gamma}_3}={\cal F}_{e,3}(\bm x), ~  {\bm x}_3={\cal F}_{d,3}(\{{\bm \gamma}_1,{\bm \gamma}_2,{\bm \gamma}_3\})\\
    \vdots\nonumber \\
    {{\bm x}_M}= {\cal F}_M({\bm x}, \{ {{\bm \gamma}_1},{{\bm \gamma}_2},\dots,{{\bm \gamma}_{M-1}} \};{\bm w}_M),~ {{\bm\gamma}_M}={\cal F}_{e,M}(\bm x), ~ {\bm x}_M={\cal F}_{d,M}(\{ {{\bm \gamma}_1},{{\bm \gamma}_2},\dots,{{\bm \gamma}_M} \}),
\end{eqnarray}
where $M$ is the number of the contained AE modes.  
{Note that, with the present hierarchical concept, we should prepare the latent vectors obtained from the encoders of the lower-order subnetworks to train the higher-order subnetworks, while training for each subnetwork is performed independently.}
With this basic model proposed here by following the original concept by \citet{SSH2004}, the number of subnetworks required to obtain up to $M$th mode is $M$, which means that a lot of AE modes are required to represent complex flow problems, e.g., turbulence.
Therefore, we also propose a concept of {\it ordered autoencoder mode family}, which contains multiple AE modes per a subnetwork.
In the present paper, the original concept, i.e., one AE mode per a subnetwork, is applied to the laminar examples, and the extended proposal with the {\it mode family} is utilized for the assessment with a $y-z$ cross-sectional velocity field of turbulent channel flow as discussed in the next section.    
Throughout the present study, a five-fold cross validation \citep{BK2019} is adopted to obtain each model.
Note in passing that we will report an ensemble averaged $L_2$ error norm over the five-fold cross validation and show a representative flow field generated from the machine-learned model which has the best validation score in the cross validation.
{In the present study, normalization and standardization are not applied to both input and output attributes because the present AE is constructed for each problem setting.}

\section{Results and Discussion}
\subsection{Example 1: Periodic wake behind a two-dimensional cylinder}

\fg{As the first example, l}et us consider a temporally periodic wake behind a circular cylinder at ${\rm Re}_D=100${, which has often been utilized as a basic example in demonstrating the ability of reduced order modeling \citep{TBDRCMSGTU2017,loiseau2018sparse,bergmann2005optimal,ehlert2019locally}}  
The training data are prepared using a two-dimensional direct numerical simulation (DNS) \citep{kor2017}. 
The governing equations are the incompressible Navier--Stokes equations, i.e.,
\begin{align}
    &\bm{\nabla} \cdot \bm{u}=0, \\
    &\dfrac{\partial\bm{u}}{\partial t} + \bm{\nabla} \cdot (\bm{uu})  = - \bm{\nabla} p + \dfrac{1}{{\rm Re}_D}\nabla ^2 \bm{u},
\end{align}
where $\bm{u}$ and $p$ denote the velocity vector and pressure, respectively.
All quantities are non-dimensionalized using the fluid density, the free-stream velocity, and the cylinder diameter.
The size of the computational domain is ($L_x, L_y$)=(25.6, 20.0), and the cylinder center is located at $(x, y)=(9,0)$.
The flow is solved on the Cartesian grid system with the grid spacing of $\Delta x=\Delta y = 0.025$, and the no-slip boundary condition on the cylinder surface is imposed by using an immersed boundary method \citep{kor2017}.
Although the number of grid points used for DNS is $(N_x, N_y)=(1024, 800)$, only the flow field around the cylinder is used as the training data whose dimension is $(N_x^*, N_y^*)$.
In the first example of periodic flow, for instance, we consider a domain of $8.2 \leq x \leq 17.8$ and $-2.4 \leq y \leq 2.4$ with $(N_x^*, N_y^*)=(384, 192)$. 
As the input and output attributes, we use the vorticity field $\omega$. 
The time interval of flow field data is 0.25 dimensionless time, which  corresponds to approximately 23 snapshots per a period, with the Strouhal number of 0.172.
Although a larger domain is considered in the second example of transient case, essentially the same computational procedure is applied.

\begin{figure}
	\begin{center}
		\includegraphics[width=0.80\textwidth]{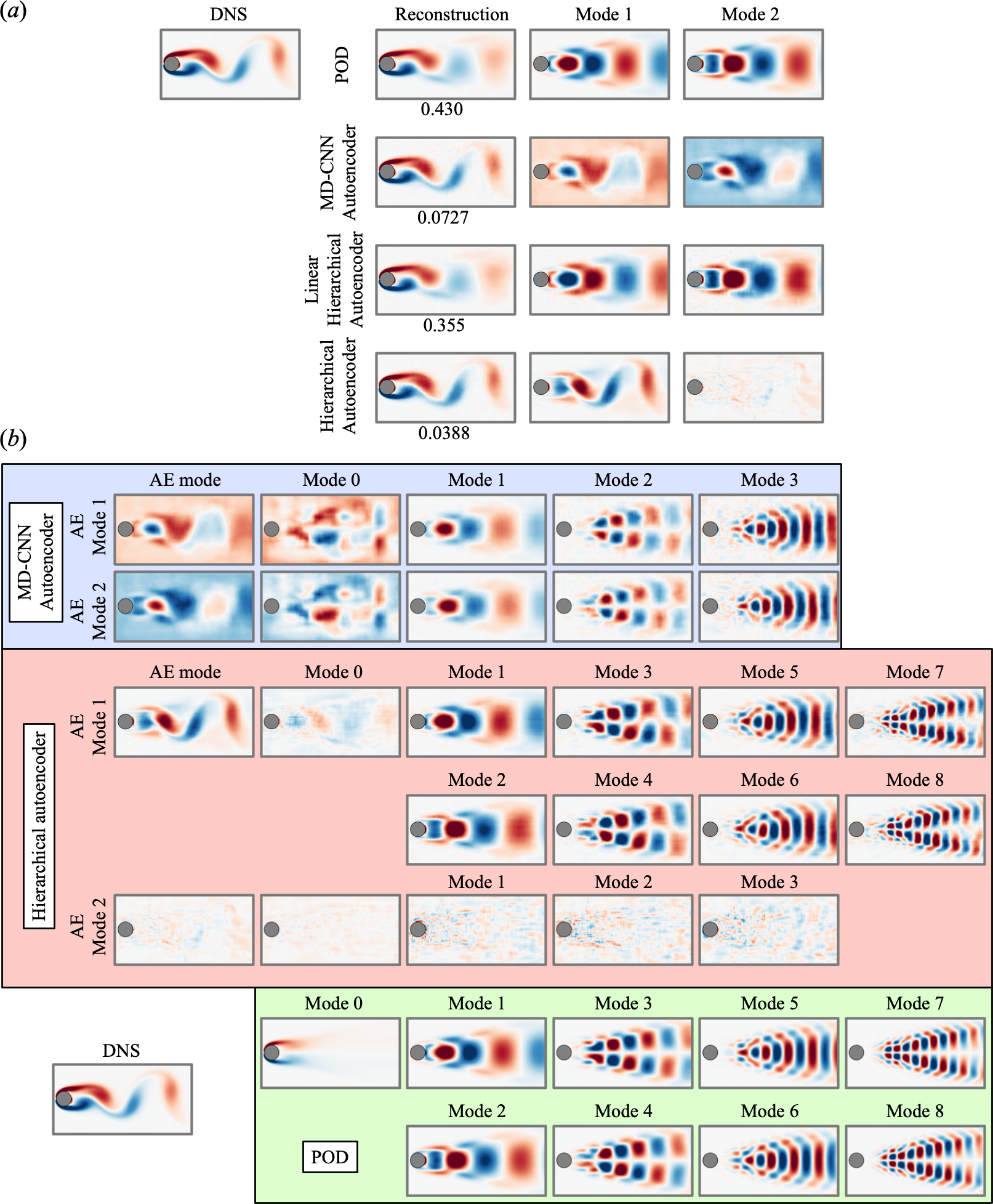}
		\caption{$(a)$ Decomposed laminar cylinder wakes.  The number of POD and AE modes are set to 2.  The values listed below the contours are the normalized $L_2$ error norm $\epsilon = ||\omega_{\rm DNS}-\omega_{\rm rec}||_2/||\omega_{\rm DNS}||_2$, where $\omega_{\rm rec}$ is the reconstructed field using the decomposed modes.  $(b)$ POD applied to AE-based modes. 
		Blue box shows the results of MD-CNN-AE proposed by \citet{MFF2019}.  Red box shows the present hierarchical AE model.  Green box shows POD modes 1 to 8 for the present DNS data.
		}
		\label{fig3}	
	\end{center}
\end{figure}

We summarize the results obtained using three types of mode decomposition methods for the periodic cylinder wake, i.e., proper orthogonal decomposition (POD), mode-decomposing convolutional neural network (MD-CNN-AE) \citep{MFF2019}, and the present hierarchical AE, in figure \ref{fig3}$(a)$.
Hereafter, the number of POD and AE modes are set to 2 in the laminar examples.
As shown here, the POD with 2 modes cannot sufficiently reconstruct the wake in terms of both the $L_2$ error norm and vorticity contour.
With the MD-CNN-AE, the reconstructed field shows good agreement with the reference DNS data thanks to the multiple POD modes contained in every single AE mode, as discussed in \citet{MFF2019}.
However, each mode obtained from the MD-CNN-AE is not sufficiently interpretable, since the AE-based modes are not orthogonal with each other and cannot be ordered following the energy containing ratio as discussed above.
Here, the hierarchical AE with linear activation is also considered to assess the ability of the hierarchical concept.
Due to the analogy between a linear AE and POD \citep{BH1989}, the linear hierarchical AE decomposes the flow into the modes akin to the POD 1 and 2 modes, as shown in the third row of figure \ref{fig3}$(a)$.
We then apply this hierarchical concept with nonlinear activation, as shown in the bottom row of figure \ref{fig3}$(a)$.
Noteworthy here is that only Mode 1 has the wake-like structure, while Mode 2 does not contain any structural information.

To further investigate these observations, let us take the POD analysis for the field reconstructed using the AE-based modes following \citet{MFF2019}.
As reported in \citet{MFF2019}, two MD-CNN-AE modes contain 6 POD mode-like structures, as shown in the blue box of figure \ref{fig3}$(b)$.  
Also, the POD modes 0 of each AE mode are decomposed in such way that the summation of these POD modes 0 are zero. 
On the other hand, as shown in the red box of figure \ref{fig3}$(b)$, with the hierarchical AE, AE mode 1 contains 8 POD mode-like structures thanks to the nonlinear activation function, although AE mode 2 contains no structural POD modes.
It suggests that the present hierarchical concept leads the machine learning models to extract the features based on their contribution for reconstruction, and this coincides {with} the fact that the periodic cylinder wake can be represented with only a scalar phase variable as revealed in the phase-reduction analysis \citep{TN2018}.  
To put it simply, just one variable is sufficient to represent this time-periodic example. 

\subsection{Example 2: Transient wake behind a two-dimensional cylinder}

\begin{figure}
	\begin{center}
		\includegraphics[width=0.95\textwidth]{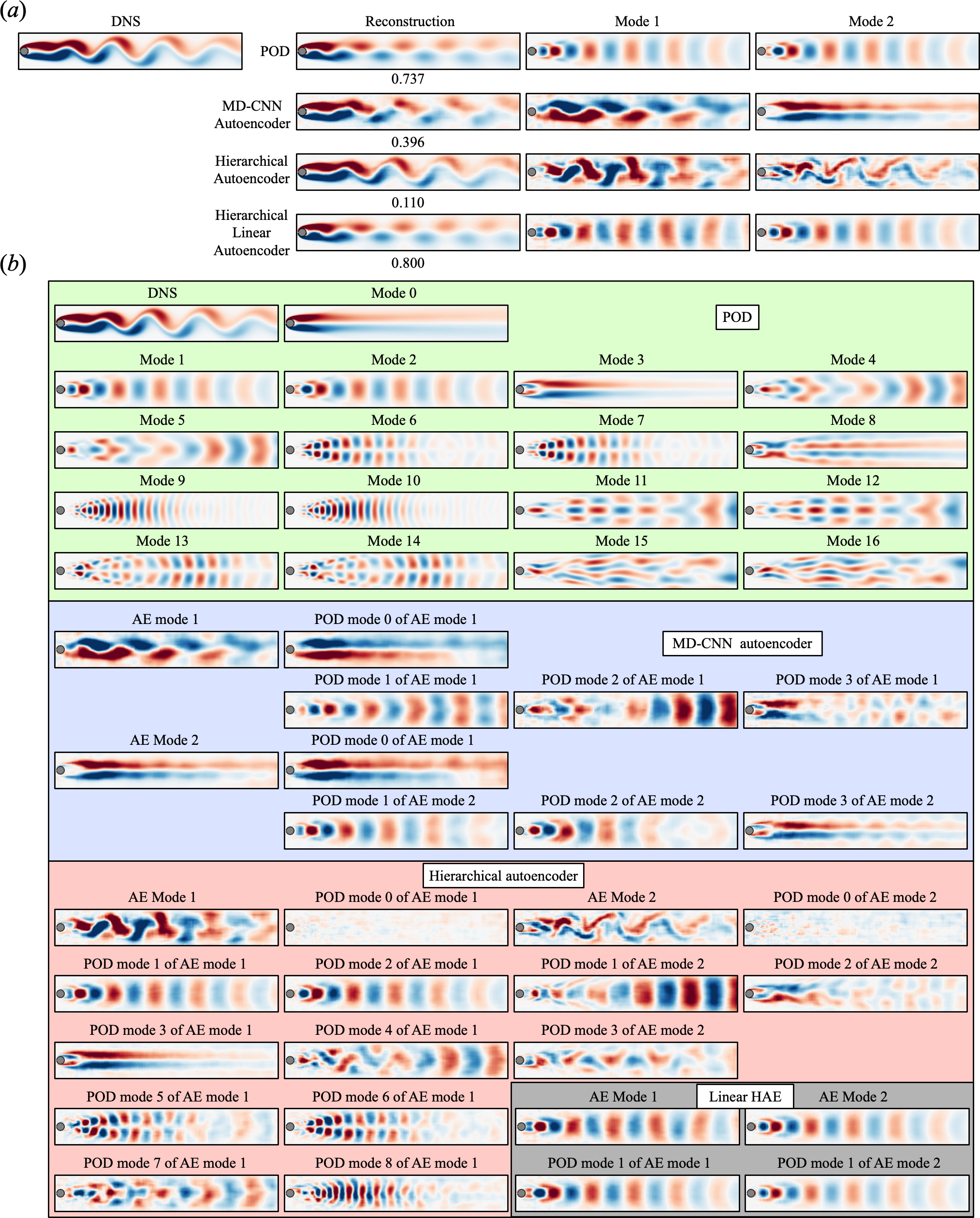}
		\caption{$(a)$ Decomposed transient wakes.  The number of POD and AE modes are set to 2.  The listed values below the contours are the normalized $L_2$ error norm $\epsilon = ||\omega_{\rm DNS}-\omega_{\rm rec}||_2/||\omega_{\rm DNS}||_2$.  $(b)$ POD applied to AE-based modes.  Blue box shows the results of MD-CNN-AE proposed by \citet{MFF2019}.  Red box shows the present model.  Green box shows POD modes 1 to 16 for DNS data.  For comparison, the results of linear hierarchical AE are also presented in gray box.}
		\label{fig4}
	\end{center}
\end{figure}

To further examine the ability of the hierarchical AE, we consider a transient wake behind the cylinder, which is a non-periodic flow.  
The training data is prepared using the DNS as explained above and extracted from the extended domain of $(L^*_x,L^*_y)=(28.8,4.8)$ with $(N^*_x,N^*_y)=(1152,192)$.

Similarly to the first example, we consider three decomposition methods as shown in figure \ref{fig4}.  
It is confirmed that only two POD modes are insufficient to reconstruct the flow field in this case, since a greater number of modes are required to reconstruct the field as compared to the first example \citep{NAMTT2003}.
On the other hand, the AE-based methods show lower $L_2$ error norm than POD thanks to the nonlinear activation function.  
Especially, the reconstructed fields with the hierarchical AE are in great agreement with the reference DNS data.  
Although AE mode 2 of the periodic cylinder wake (i.e., example 1) contained no structural information as shown in figure \ref{fig3}$(a)$, AE mode 2 in this case has some structures which cannot be extracted by the first subnetwork ${\cal F}_1$.  

Next, we also apply POD to these AE-based reconstructed fields as shown in figure \ref{fig4}$(b)$.
With the MD-CNN-AE, POD mode-like structure can be observed in the decomposed field although the modes are not properly ordered unlike the POD, and high-order modes are perhaps blended, e.g., POD mode 3 of AE mode 1 in the blue box.   
On the other hand, with the hierarchical AE, the AE mode 1 clearly contains the dominant POD modes as shown in figure \ref{fig4}$(b)$.  
Especially, what is striking here is that the POD results for the AE-based mode 2 has various blended structures of POD modes as shown in the right part of the red box in figure \ref{fig4}$(b)$.  
Although clear POD-like modes cannot be obtained from the AE-based mode 2 here because of the nonlinearity given as the activation function in the AE structure, the AE-based modes can be ordered through the concept of the hierarchical AE.  
For comparison, let us briefly touch the result of hierarchical AE with linear activation as shown in figures \ref{fig4}$(a)$ and $(b)$.  
In this case, the linear AE can extract only the fields akin to POD modes 1 and 2 due to the analogy between the linear AE and POD \citep{BH1989}, although the order follows their energy contribution.  
This comparison enables us to check the ability of hierarchical concept and the strength of the nonlinear activation of the AE again.

\subsection{Example 3: $y-z$ cross-section of turbulent channel flow}

As an example application to turbulent flow fields, let us consider the use of a $y-z$ cross-sectional velocity field of turbulent channel flow at ${\rm Re}_{\tau}=180$ under a constant pressure gradient condition. 
{Note again that the spatial order reduction of turbulence is a challenging problem with conventional ROM tools since a vast range of scales is contained in flow fields \citep{alfonsi2006,MPMF2019}.}
The training data set are obtained by direct numerical simulation \citep{FKK2006}{, which has been validated by comparison with spectral DNS data of \citet{moser1999direct}}.
The governing equations are the incompressible Navier--Stokes equations,
\begin{align} 
&\bm{\nabla} \cdot {\bm u} = 0\\
&{ \dfrac{\partial {\bm u}}{\partial t}  + \bm{\nabla} \cdot ({\bm u \bm u}) =  -\bm{\nabla} p  + \dfrac{1}{{\rm Re}_\tau}\nabla^2 {\bm u}},
\end{align}
where $\displaystyle{{\bm u} = [u~v~w]^{\mathrm T}}$ represents the velocity with $u$, $v$ and $w$ in the streamwise ($x$), wall-normal ($y$) and spanwise ($z$) directions.  
Here, $t$ is the time, $p$ is the pressure, and ${{\rm Re}_\tau = u_\tau  \delta/\nu}$ is the friction Reynolds number.  
The quantities are non-dimensionalized with the channel half-width $\delta$ and the friction velocity $u_\tau$.  
The size of the computational domain and the number of grid points here are $(L_{x}, L_{y}, L_{z}) = (4\pi\delta, 2\delta, 2\pi\delta)$ and $(N_{x}, N_{y}, N_{z}) = (256, 96, 256)$, respectively.  
{The spatial discretization is performed using the energy-conserving second order finite difference scheme on the staggered grid \citep{kajishima1999finite,ham2002fully}.
The temporal integration is conducted with the low-storage, third-order Runge-Kutta/Crank--Nicolson
scheme \citep{spalart1991spectral} and the high-order SMAC-like velocity-pressure coupling scheme \citep{dukowicz1992approximate}. 
The pressure Poisson equation is solved utilizing the fast Fourier transform in the $x$ and $z$ directions and the tridiagonal
matrix algorithm in the $y$ direction.}
The grids in $x$ and $z$ directions are set to uniform and non-uniform grid is utilized in $y$ direction.  
No-slip boundary condition is imposed on the walls and the periodic boundary condition is used in $x$ and $z$ directions.  
In the present paper, we use the fluctuation components of a $y-z$ cross-sectional velocity as the input and output attributes for the construction of AEs.

\begin{figure}
	\begin{center}
		\includegraphics[width=0.80\textwidth]{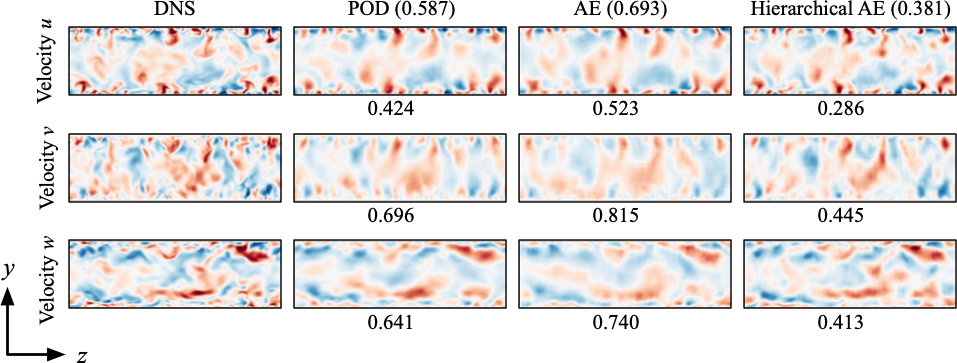}
		\caption{Reconstructed velocities from POD, conventional autoencoder, and hierarchical autoencoder at the number of the latent space $n_\gamma=288$.  Listed values are $L_2$ error norm $\epsilon$. }
		\label{fig5}
	\end{center}
\end{figure}

\begin{figure}
	\begin{center}
		\includegraphics[width=1.00\textwidth]{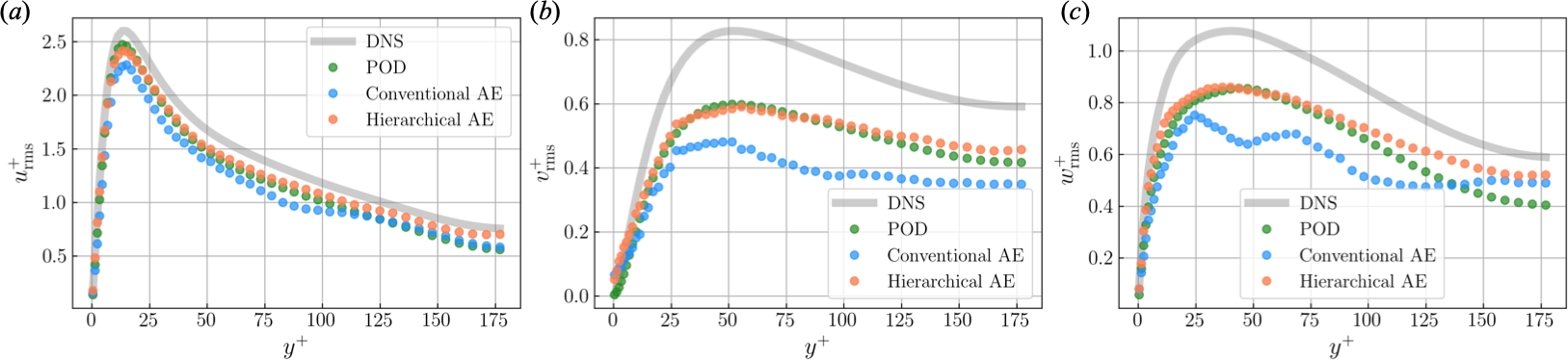}
		\caption{Root mean square value of velocity fluctuation components. $(a)$ Streamwise, $(b)$ wall-normal, and $(c)$ spanwise velocity.} 
		\label{fig6}
	\end{center}
\end{figure}

As introduced in section \ref{sec2}, hereafter we use the concept of {\it ordered autoencoder mode family} (AE mode family) to handle a flow field where a lot of modes are required to reconstruct the field, i.e., turbulence.  
In the present paper, four mode families are utilized such that each subnetwork maps the high dimensional fields into $M/4$ dimensions when $M$ modes are contained in total.
{Note in passing that although we select 4 as the number of families, a proper choice may be required to achieve a reasonable compression since the results of the hierarchical mode decomposition is affected by the results of subnetworks for low-order modes because of the hierarchical manner.}
As the first comparison of various low-dimensional mapping methods, let us present in figure \ref{fig5} the reconstructed velocity contours.
Here, we set the number of latent vector $n_{\gamma}$ to be 288 $(=72~{\rm modes/family} \times 4~{\rm families})$.  
Note that we do not use the MD-CNN-AE for this assessment of turbulence because our interest here is whether the hierarchical concept enables us to present efficient compression while keeping information of high-dimensional flows or not.
As shown, significant differences cannot be observed in the comparison between POD and conventional AE, despite the nonlinearity of AE model.
On the other hand, the $L_2$ error norm of the hierarchical AE shows smaller values than other two methods and the finer scale can also be reconstructed as shown in figure \ref{fig5}. 
It suggests that the hierarchical concept enables AE models to present efficient compression and this observation is analogous to previous reports \cite{SSH2004,saegusa2005nonlinear}.
These trends can also be observed in the statistical assessment in figure \ref{fig6} as explained later.
{In addition, notable here is that the better reconstruction of $u'$ can be seen than that of $v'$ and $w'$ for all schemes.
This is likely due to the fact that the streamwise velocity has dominant features in a turbulent channel flow. 
}

\begin{figure}
	\begin{center}
		\includegraphics[width=1.00\textwidth]{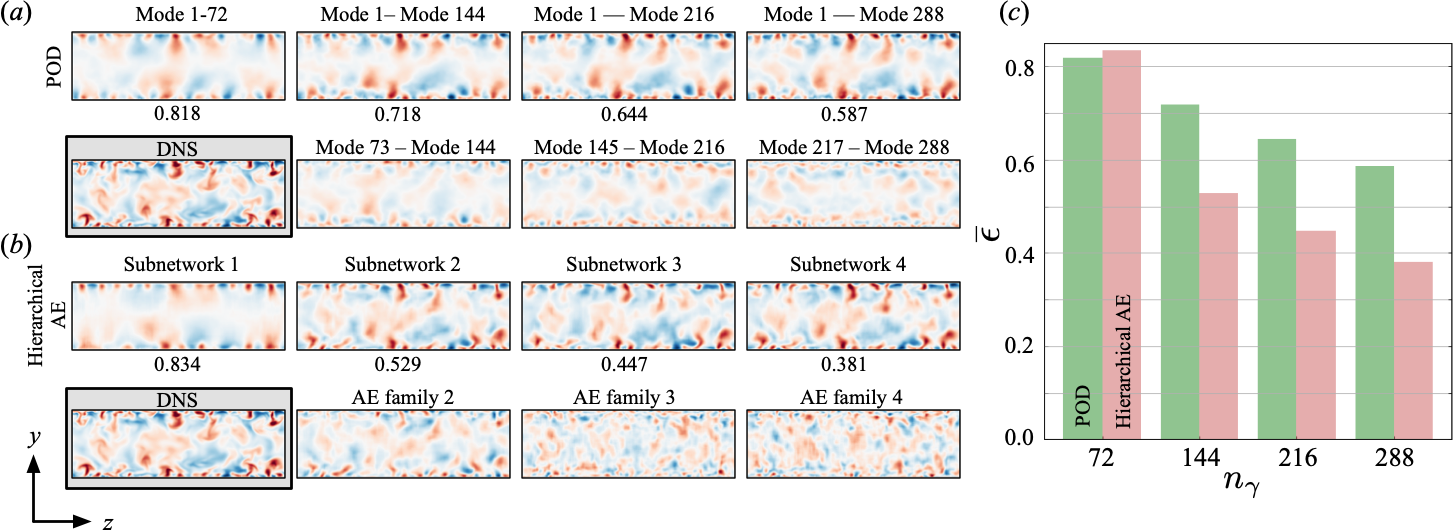}
		\caption{Summary of the contribution of each POD mode or AE mode family for the flow reconstruction at $n_\gamma = 288$.  The color contours are streamwise velocity: $(a)$ POD; $(b)$ Hierarchical autoencoder.  With both plots, the first rows indicate the summation of modes or mode families.  The second rows show each mode family. The reference DNS instantaneous field is also presented for comparison.  Listed values are ensemble $L_2$ error norm of three-velocity components, which are also presented in $(c)$. } 
		\label{fig7}
	\end{center}
\end{figure}

We also assess the ability of the hierarchical AE with a root-mean-square (RMS) value of the velocity fluctuation as shown in figure \ref{fig6}.  
Here, we also set the number of the dimension of the latent vector to be $n_\gamma = 288$. 
As shown here, the hierarchical AE shows better performance than the POD and conventional AE.
Noteworthy here is that the distributions obtained from the conventional AE show the non-smooth curve, i.e., especially $v$ and $w$, against the POD and hierarchical model which have the ordered-mode concept.
This observation also suggests that the model can obtain the nonlinear feature of the data to be aided by the hierarchical-based learning process with nonlinear activation.
{It is also striking that the curves in figure \ref{fig6}$(b)$ of both the conventional and the hierarchical AE present a sudden but slight change in the trend roughly at $y^+ \approx 25$.
This observation can also be found with the conventional AE as shown in figure 6$(b)$.
This is likely due to the fact that it is not necessary to satisfy the continuity in results generated by machine learning models, while POD bases are able to guarantee this point.
It may be mitigated by considering a physics manner in a loss function \citep{LY2019,raissi2020hidden}, although it remains in future.}

Next, let us examine in figure \ref{fig7} the contribution of each mode obtained from POD and hierarchical AE for the flow reconstruction with $n_{\gamma}=288$.  
With both cases, the $L_2$ error norms listed below the color contours decrease with increasing the number of modes contained in the latent space.  
As a whole, the hierarchical AE shows its great advantage against the POD reconstruction in terms of the $L_2$ error norm and the velocity color contours, although POD outperforms the first subnetwork which is the same as the conventional AE with $n_{\gamma}=72$.  
It is striking that high-order AE-mode families have finer structure not only in the near-wall region but also in the channel center despite that the high-order modes with the POD represent the fine structure only near the wall as shown in figure \ref{fig7}.  
This is likely due to the nonlinear activation in the machine learning model.  
In addition, we also find that the lower AE mode families obtained by the subnetworks 1 and 2 extract the large scale structure thanks to the hierarchical manner of the proposed method.

\begin{table}
\begin{center}
\def~{\hphantom{0}}
\begin{tabular}{cccccc}\hline\hline
     		   & POD & Conventional AE & Hierarchical AE \\\hline
     Training [{\rm min.}] & 4.43 & 571.5 & 400.8\\
     Reconstruction [{\rm s}] & 7.14&$3.13\times10^{-3}$&$3.21\times10^{-3}$ \\\hline\hline
    \end{tabular}
    \\
  \caption{{Computation time for training the machine learning models, calculating (noted as training in the table) the POD bases, and reconstructing 1 snapshot of turbulent channel flow example at $n_{\gamma}=288$. The POD is performed using CPU (Xeon Gold 6130, 2.1 GHz) and the AEs are constructed with GPU (NVIDIA Tesla V100).}}
  \label{tab1}
\end{center}
\end{table}

{The comparison of computational cost among POD, conventional AE, and hierarchical AE is summarized in table \ref{tab1}.
The POD is performed using CPU (Xeon Gold 6130, 2.1 GHz) and the AEs are constructed with GPU (NVIDIA Tesla V100).
In this demonstration, we use 10000 snapshots as training data for all schemes.
To derive the conventional and hierarchical AE models, approximately 10 h and 7 h, respectively are taken.
Note here that the AE models can provide better compression with faster reconstruction time than POD once we train the models.
}

\begin{figure}
	\begin{center}
		\includegraphics[width=1.00\textwidth]{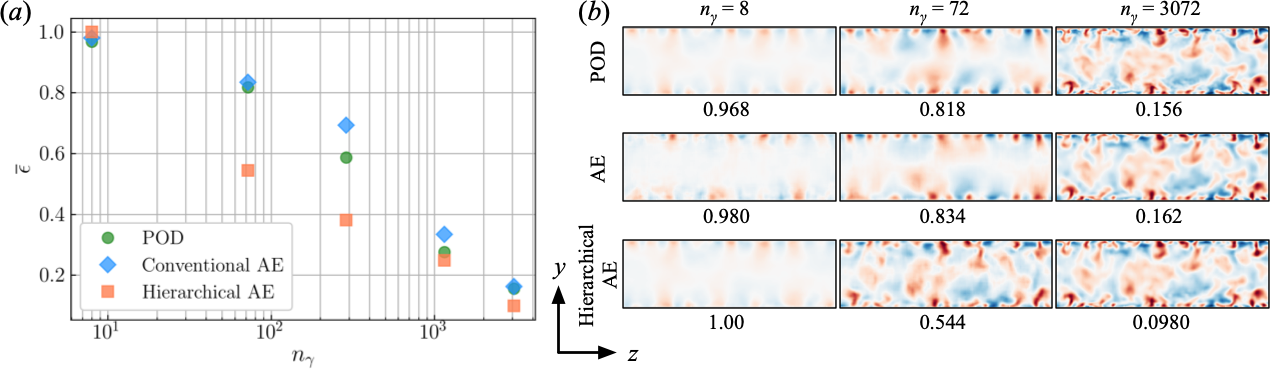}
		\caption{Dependence on the number of the latent space $n_\gamma$.  $(a)$ $L_2$ error norm $\epsilon$ $(b)$ The streamwise velocity contours of three decomposition methods at $n_\gamma=8,72,$ and 3072. Listed values are ensemble $L_2$ error norm of three-velocity components, which are the same as those presented in $(a)$.}
		\label{fig8}
	\end{center}
\end{figure}

At last, we investigate the dependence on the number of the latent space $n_\gamma$, as summarized in figure \ref{fig8}. 
As shown in figure \ref{fig8}$(a)$, the hierarchical AE outperforms POD and conventional AE in terms of the $L_2$ error norm for the entire range of $n_\gamma$.
The exception is the case of $n_\gamma=8$, at which none of the decomposition methods can reconstruct the flow field well due to the excessive low dimensionalization.  
In particular, the strength of the hierarchical concept can be seen at $n_\gamma=72$.  
Obviously, the finer scale can also be caught by the hierarchical AE despite the low fidelity of the other two methods.  
Since the POD and conventional AE are also able to reconstruct the flow field well, the major difference cannot be observed with the larger $n_\gamma$, i.e., $n_\gamma=3072$.  
Through the examination with the turbulent channel flow, we see the great potential of the concept of the hierarchical AE with mode family as the new augmented nonlinear mode decomposition method.

\section{Conclusions}

We proposed a new nonlinear mode decomposition method with convolutional neural network (CNN) inspired by the hierarchical autoencoder (AE) which can order the AE modes following their contributions to the reconstructed field while achieving efficient order reduction.  
The proposed concept was first applied to a laminar cylinder wake and its transient process as preliminary tests. 
It was revealed that the first AE mode contains dominant structures that appear in the high-order proper orthogonal decomposition modes 
when the number of latent vectors is set to be 2.  
The method was further investigated with a $y-z$ cross-sectional velocity of turbulent channel flow at ${\rm Re}_{\tau}=180$ as an example application to turbulent flows.  
We also proposed the concept of {\it mode family}, which groups some AE modes in the latent space per a subnetwork, so that we can handle more complex flows where the high-order spatial modes are required to reconstruct the flow field.  
We found that the hierarchical AE with the concept of the {\it mode family} has a strong potential as the nonlinear low-dimensional mapping function.

Although we revealed the ability of the new nonlinear mode decomposition method with the hierarchical concept, some issues are remaining here: one of them is the non-uniqueness.  
Because the training process of neural networks is probabilistic, the results of each process are not unique even if a cross-validation is performed well.  
It suggests that some advanced designs, which can account for the probabilistic view into AE models, are desired \cite{MFRFT2020}.
Otherwise, due to the hierarchical structure, the results of the hierarchical mode decomposition with machine learning is affected by the results with subnetworks for low-order modes.  
Although it is still difficult to avoid these issues with the probabilistic manner, we strongly believe that the nonlinear mode decomposition with the hierarchical concept can allow us to examine the complex flow phenomena efficiently.  
For instance, the proposed method, which can map a high-dimensional system into a low-dimensional space with the hierarchical sense, may be useful for reduced order modeling of turbulent flow fields requiring a lot of spatial modes.
{\citet{SGASV2019} have recently demonstrated the possibility of long short-term memory (LSTM) based temporal prediction for turbulence considering the nine-equation turbulent shear flow model. 
We can consider to combine the extracted hierarchical modes with the LSTM, which is analogous idea to \citet{HFMF2020a}.}
{The present idea can also be extended to three-dimensional flows by replacing two-dimensional operations as three-dimensional functions in the CNN.}
{For applications to 
higher Reynolds number turbulence, the combination of hierarchical concept and generative adversarial networks, which have recently been utilized to enhance fine scale information of turbulence \citep{KKWL2020}, may also be considered.}
In addition, the combination with the operator-driven method, i.e., resolvent analysis \citep{MS2010,abreu2020spod}, can also be expected.
Usually, these methods can be utilized since we know a contribution of each mode for energy ratio as eigen- or singular values.  
{We may be able to consider uses of customized POD, e.g., multi-scale POD \citep{mendez2019multi} and spectral POD \citep{towne2018spectral,abreu2020spod}, to augment data compressionability.
Although we suspect whether these sophisticated matrix factorizations work well or not highly depends on target flows, we can strongly expect the combination with them.}
Although these expectations are just example, we hope that the present concepts will help the field of fluid dynamics.

\section*{Acknowledgements}
This work was supported through JSPS KAKENHI Grant Number 18H03758.  
We acknowledge Mr. Takaaki Murata, Mr. Masaki Morimoto, and Mr. Naoki Moriya (Keio Univ.), and Mr. Kazuto Hasegawa (Keio Univ./Polimi) for stimulating discussions and fruitful comments.

\section*{Data Availability}
The data that support the findings of this study are available from the corresponding author upon reasonable request.

\section*{References}


\bibliography{FNF2020}
\end{document}